\documentclass[3p, number, twocolumn]{elsarticle}
\usepackage[utf8x]{inputenc}
\usepackage{graphicx}   

\begin{document}

\begin{frontmatter}
\title{A new solution for mirror coating in $\gamma$-ray Cherenkov Astronomy}

\author[unituebingen]{A.~Bonardi\corref{cor1}}
\ead{antonio.bonardi@uni-tuebingen.de}

\author[unituebingen]{G.~P\"uhlhofer}
\author[unituebingen]{S.~Hermanutz}
\author[unituebingen]{A.~Santangelo}
\cortext[cor1]{Corresponding author}
\address[unituebingen]{University of T\"ubingen, Institut f\"ur Astronomie und Astrophysik T\"ubingen,\\ Sand 1, 72076 T\"ubingen, Germany}

\begin{abstract}
In the $\gamma$-ray Cherenkov Astronomy framework mirror coating plays a crucial role in defining the light response of the telescope. We carried out a study for new mirror coating solutions with both a numerical simulation software and a vacuum chamber for small sample production.
In this article, we present a new mirror coating solution consisting of a 28-layer interferometric SiO$_{2}$-TiO$_{2}$-HfO$_{2}$ design deposited on a glass substrate, whose average reflectance is above $90\%$ for normally incident light in the wavelength range between 300 and 550~nm. 
\end{abstract}
\end{frontmatter}

\section{Introduction}

In $\gamma$-ray Cherenkov Astronomy mirror coating plays a crucial role in defining the optical properties of the Imaging Atmospheric Cherenkov Telescopes (hereafter IACTs) and, hence, their sensitivity. 
The Cherenkov light produced by a $\gamma$-ray induced atmospheric shower is reflected and focused towards the focal plane by the reflector, which can be a single dish (Davies-Cotton \cite{davies_cotton}, parabolic \cite{magic_telescope},\cite{hessII_telescope} , or intermediate design \cite{mst_mechanics}) or composed of a primary and a secondary dish (Schwarzschild-Couder design \cite{vassiliev2007}). The Cherenkov light is then collected by the IACT camera photo-detectors, and the event is recorded.\\

\subsection*{The Night Sky Background}
\begin{figure}[!htb]
  \centering
  \includegraphics[width=0.4\textwidth]{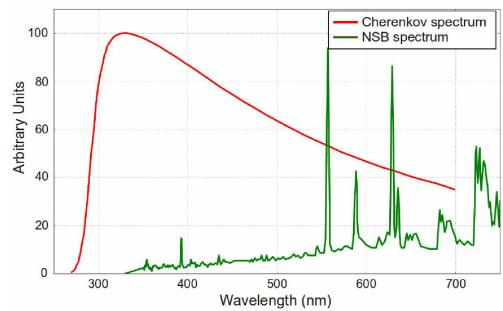}
  \caption[cherenkov_spectrum]{\footnotesize Cherenkov light and NSB spectrum in La Palma (Canary Islands, Spain) at 2200~m~a.s.l., arbitrary units~\cite{benn}.}
\label{cherenkov_spectrum}
\end{figure}

The light collected by the IACT camera contains a large fraction of background due to solar, stellar and moon (in case of observations carried out during moonlight) diffuse light in the Earth atmosphere, whose total contribution is called Night Sky Background (NSB).
In fig. \ref{cherenkov_spectrum} the spectrum of the emitted Cherenkov light is shown together with the NSB spectrum for a moonless night in La Palma (Canary Islands, Spain) as reported by Benn and Ellison \cite{benn}.
It is therefore advisable to tune the reflectance so to maximize the collection of Cherenkov photons and, at the same time, to reduce the amount of NSB collected photons. \\

\subsection*{Present IACT mirrors}
\label{Present_IACT_mirrors}
Present generation of IACT reflectors are composed of mirror tiles with either a solid glass or an Aluminum honeycomb structure. Except for MAGIC I and large part of MAGIC II reflector, the mirror tiles have a front-coating (i.e the reflective layer is on the mirror tile side that faces the telescope camera) over a glass substrate. The glass substrate is either the mirror structure itself (H.E.S.S. and VERITAS telescopes \cite{hess_optics},\cite{veritas_telescope}) or an additional glass sheet over the honeycomb structure (``cold slumping technique'', used for part of the MAGIC II reflector \cite{coldslumping}). At present time, the mirror cold slumping technique is the most investigated technology for the CTA project, due to the combination of reduced cost and light weight, even if alternative mirror manufacturing techniques are investigated as well \cite{pareschi},\cite{afoerster747}.\\ 
For almost all the cases of mirror coating on a glass substrate, the coating consists of an Aluminum reflective layer plus a protective layer made of SiO$_{2}$ (the H.E.S.S. telescopes \cite{hess_optics}, and part of the MAGIC II telescopes \cite{coldslumping}), Al$_{2}$O$_{3}$ (the four VERITAS Telescopes \cite{veritas_telescope}), or an interferometric 3-layer SiO$_{2}$-HfO$_{2}$-SiO$_{2}$ (used recently for the mirror refurbishment on three of the four H.E.S.S. I telescopes). In the latter case, the protective layer enhances also the mirror reflectance by about $5\%$. 
Apart from Aluminum based coatings, a dielectric coating with a reflectance~$>~95\%$ based on different materials has been developed for H.E.S.S. and CTA (Cherenkov Telescope Array) \cite{afoerster936},\cite{afoerster755} and applied to 99 H.E.S.S. mirrors for a longterm test in real environmental conditions.
In the particular case of the MAGIC I and large part of MAGIC II reflector, the Aluminum honeycomb structure is diamond milled, no further reflective coating is needed, and only a protective Al$_{2}$O$_{3}$ layer is applied \cite{doro2008}.\\
The main advantage of using Aluminum as reflective layer is its high reflectance ($\sim90\%$) all over the Cherenkov spectrum. On the other hand, for the same reason, no NSB suppression is achieved. A further drawback is the mirror lifetime which is limited to few years. Since IACTs are usually in severe environments without any protective dome, Aluminum oxidation and detachment from the substrate may occur, leading to a general reflectance drop.\\


\subsection*{Interferometric mirror coating}

As stated above, interferometric dielectric multi-layer designs are an alternative to Aluminum based coatings. When light passes from a dielectric material to another one, it is reflected according to Fresnel's law. The reflectance for small incident angles is equal to

\begin{displaymath}
\qquad \qquad \quad R = \left(\frac{n_{1}-n_{2}}{n_{1}+n_{2}}\right)^{2}
\end{displaymath}

where $R$ is the reflection coefficient, and $n_{1}$ and $n_{2}$ the refractive indexes of the first and second material, respectively.\\

In a multi-layer coating design, the reflectance value depends also on the wavelength of the incident light because of the interferometric principle.
For a light ray passing from one layer to a second one, the highest reflectance value is obtained for

\begin{displaymath}
\qquad \qquad \qquad \lambda = \frac{4\cdot n_{1} \cdot L_{1}}{\cos(\theta)}
\end{displaymath}

where $\lambda$ is the incident light wavelength in vacuum, $\theta$ the incident angle, and $L_{1}$ and $n_{1}$ the thickness and the refractive index of the first layer, respectively.
By alternating several dielectric layers with different thickness and refraction index, it is possible to achieve very high reflectance within the desired wavelength range and, at the same time, very low reflectance outside of it. Furthermore, since no metallic layer is used, no deterioration due to oxidation will occur, and then the mirror lifetime is expected to be longer.\\

Dielectric coatings may face however challenges, if applied to Cherenkov astronomy. In fact, a large number of layers ($> 25$) is required, which is a cost and technological challenge for mass production. The large number of layers increases the global manufacturing time, which is the leading component of the total mirror coating cost. Furthermore, many promising materials need particular treatments during or after the deposition, ending in a further cost increase. In the case of the 99 H.E.S.S. mirrors with dielectric coating, the cost increases with respect to the standard Aluminum based coating of about~$30\%$. 
It has also been pointed out that  the large number of layers could result in an accelerated layer deterioration due to the thermally induced stresses between different layers. No evidence of such phenomenon has been observed on the H.E.S.S. dielectric mirrors, but it cannot be \emph{a priori} excluded for any other interferometric coating solution. Furthermore, a higher humidity condensation rate has been observed compared to metallic coatings \cite{bib:pchadwick}.\\

\section{The coating study}

The purpose of our work was to investigate new front-coated interferometric coating solutions on glass substrate, which could be easily applied to all the glass-substrate-based mirrors of the future CTA IACTs, as described in the CTA concept document \cite{CTA_concept}. The potential problems involved with interferometric coatings described above were not part of this work and, hence, further investigations on those issues will be needed.\\

In this study we set the following requirements for our interferometric coating solutions:

\begin{itemize}
 \item [a)] for normally incident light, reflectance above $90~\%$ within the wavelength range between 300 and 550 nm (hereafter WR$_{300-550}$) and very low reflectance at longer wavelengths. Thus, a substantial fraction of Cherenkov photons is reflected to the focal plane, while the NSB is strongly suppressed (see fig.~\ref{cherenkov_spectrum});
 \item [b)] no mirror thermal treatment allowed during or after the manufacturing. In this way, the coating solution is also suitable not only for glass made mirrors, but for mirrors with a honeycomb structure as well;
 \item [c)] technological easiness: the mirror coating procedure has to be as simple as possible to reduce the production cost and to permit, in principle, the mirror coating and re-coating on IACT site (logistics cost suppression).
\end{itemize}

We selected SiO$_{2}$ as low refractive index material, because of its easiness in evaporation and deposition, good transparency, and good availability on the market. We chose TiO$_{2}$ as high refractive index material, because of its hardness and chemical stability. Similarly to what reported by Duyar et al. \cite{bib:duyar}, TiO$_{2}$ layers were deposited through Ti$_{3}$O$_{5}$ reactive evaporation in an O$_{2}$ atmosphere. The reason of using Ti$_{3}$O$_{5}$ as starting material (instead of directly TiO$_{2}$) is that Ti$_{3}$O$_{5}$ can evaporate only as TiO (Titanium monoxide) with stable stoichiometric ratio. On the contrary, TiO$_{2}$ can produce vapors with different Ti-O stoichiometric ratios. Usually this ratio changes during the evaporation, resulting in a different refractive index for every TiO$_{2}$ layer, which makes TiO$_{2}$ not suitable for mass production \cite{bib:duyar}.\\

\subsection*{Instrumentation and Procedure}

\begin{figure}[!htb]
  \centering
  \includegraphics[width=0.4\textwidth]{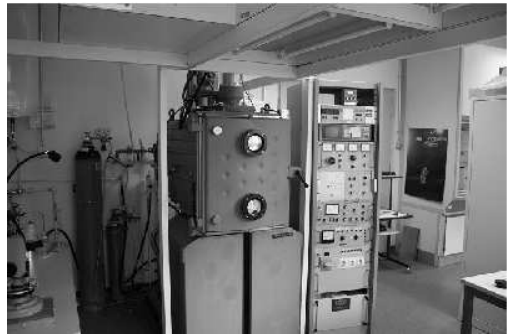}
  \caption[vacuum_chamber]{\footnotesize The Balzers BA k 550 vacuum chamber at the IAAT.}
\label{vacuum_chamber}
\end{figure}

The work presented here involves two stages. In the first stage, we used the McLeod Concise \cite{McLeod} commercial simulation software for investigating different coating solutions and optimizing their designs. The second stage consisted of preparing coating samples on small glass substrates (5~x~5~cm$^{2}$) in the Balzers BA k 550 vacuum chamber based at the Institut f\"ur Astronomie and Astrophysik T\"ubingen (IAAT), and shown in Fig.~\ref{vacuum_chamber}. The reflectance of the produced samples was then compared with simulation expectations, in order to validate or reject the coating solution.\\

Inside the vacuum chamber a residual vacuum pressure $\sim~10^{-6}$~mbar is achieved within two hours by a pneumatic and turbo-molecular pumping system. Dielectric coating materials are thermally evaporated in a Leybold ESV6 4-fold electron beam crucible, and the thickness of the deposited materials is monitored by a 6-fold Provac QSP 650 quartz micro-balance installed in the center of the chamber ceiling. Samples are fixed to a rotating wheel, so the same amount of material is deposited on all the prepared samples. The chamber is equipped with a Nitrogen and Oxygen flooding system for humidity removal and oxide active deposition respectively. Furthermore, a water warming/cooling system ($15-65~^{\textrm o}$C) is present. It is used both for cooling down the whole chamber during material evaporation and for preventing moisture formation inside the chamber after it has been opened.\\ 

\subsection*{Preliminary Study}

Since we used neither any substrate heating system nor any post-production thermal annealing, using our coating setup we expected to obtain TiO$_{2}$ layers with smaller mass density (and hence refractive index) than the bulk material ($\rho=3.84~\mathrm{g}~\mathrm{cm}^{-3}$, for TiO$_{2}$ anatase), as also reported in similar circumstances by Jerman and Mergel \cite{bib:jerman}. Therefore, the first step was to determine the real refractive index of the deposited TiO$_{2}$  layers. We produced some preliminary 4-layer TiO$_{2}$-SiO$_{2}$ coating samples using the same experimental setup as used afterwards (P(O$_{2}$)~=~$10^{-4}$ mbar, TiO$_{2}$ deposition rate $\sim~0.1~\mathrm{nm}~\mathrm{s}^{-1}$, substrate temperature 15 $^{\textrm o}$C, no thermal annealing). By comparing the observed reflectance curve with the simulated one, we observed that the refractive index of the deposited TiO$_{2}$ was $5\%$ smaller than the value provided by the McLeod Concise software, which was referring to the bulk material. Therefore, we modified the TiO$_{2}$ refractive index in the simulation software according to this result.\\ 

\section{Results}

\begin{figure}[!htb]
  \centering
  \includegraphics[width=0.4\textwidth]{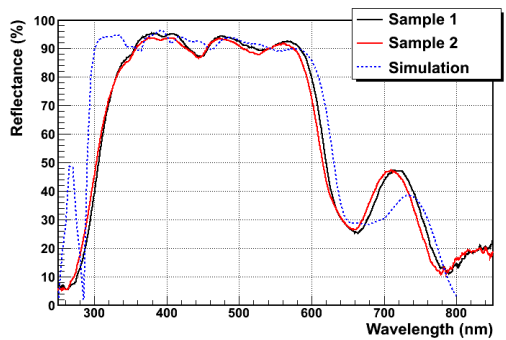}
\caption[TiO2-SiO2_24layers]{\footnotesize Measured reflectance vs. wavelength for normally incident light on a 24-layers TiO$_{2}$-SiO$_{2}$ coated glass substrate (red and black line) and simulation (blue dashed line). The reflectance was measured with an Ocean Optics JAZ hand reflectometer, kindly provided by MPIK Heidelberg.}
\label{TiO2-SiO2_24layers}
\end{figure}

The next step was to produce a 24-layer TiO$_{2}$-SiO$_{2}$ design which, according to the McLeod Concise simulation software, was able to provide $\sim~90\%$ reflectance in the whole WR$_{300-550}$. The manufactured samples showed a much worse reflectance in the wavelength range $<350$ nm than what was expected from the simulation, as shown in Fig.~\ref{TiO2-SiO2_24layers}.

\begin{figure}[!htb]
  \centering
  \includegraphics[width=0.4\textwidth]{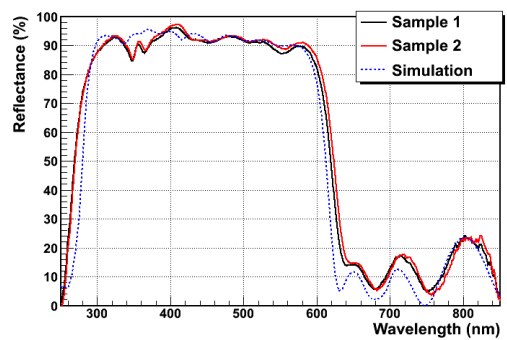}
\caption[Reflectance_TiO2-SiO2-HfO2_28layers]{\footnotesize Measured reflectance vs. wavelength for normally incident light on a 24-layers TiO$_{2}$-SiO$_{2}$ plus 4-layers HfO$_{2}$-SiO$_{2}$ coated glass substrate (red and black line) and simulation (blue dashed line). The reflectance was measured with an Ocean Optics JAZ hand reflectometer, kindly provided by MPIK Heidelberg.}
\label{Reflectance_TiO2-SiO2-HfO2_28layers}
\end{figure}

The reflectance discrepancy between simulation and manufactured samples was found to be due to the
extinction coefficient of the TiO$_{2}$ layers, which was significantly underestimated by the simulation software for wavelengths below 350 nm. We modified the TiO$_{2}$ extinction coefficient to match the experimental results. Unfortunately, it turned out it was not possible to obtain the required $\sim90\%$ reflectance in WR$_{300-550}$ by using only TiO$_{2}$-SiO$_{2}$. As a consequence, we created a new design consisting of 24 alternating TiO$_{2}$-SiO$_{2}$ layers plus 4 alternating HfO$_{2}$-SiO$_{2}$ layers on the top, in order to enhance the reflectance in the 300-350 nm region. We chose HfO$_{2}$ among other high refractive index materials because of its excellent transparency in the ultraviolet region, its hardness, its chemical stability, and because it does not need any thermal treatment, like TiO$_{2}$.\\ 
The reflectance spectrum for normally incident light of two samples of the new 28-layer TiO$_{2}$-SiO$_{2}$-HfO$_{2}$ design produced in the IAAT coating chamber is shown in Fig.~\ref{Reflectance_TiO2-SiO2-HfO2_28layers}, together with the simulated spectrum.\\ 

The reflectance curves presented in Fig.~\ref{Reflectance_TiO2-SiO2-HfO2_28layers} show a very good agreement between the simulation prediction and the manufactured samples. The only significant discrepancy in WR$_{300-550}$ is for wavelengths between 340 and 380 nm, where the experimental sample reflectance is $\sim5\%$ lower than the simulated one. That could be due either to the uncertainty in the deposition thickness determination of one or more layers or to the above stated underestimation of the TiO$_{2}$ extinction coefficient. Beyond the WR$_{300-550}$, the reflection curve of the manufactured samples shows smother edges compared to the simulation, and a slightly higher reflectance between 600 and 750 nm. That can be explained by inhomogeneities between two consecutive manufactured layers, while the simulation software assumes perfectly plane layers. Anyhow, the average reflectance in WR$_{300-550}$ of our manufactured samples is $91.3\%$ (sample 1) and $92.0\%$ (sample 2), respectively, well above the required $90\%$.\\

\begin{figure}[!htb]
  \centering
  \includegraphics[width=0.4\textwidth]{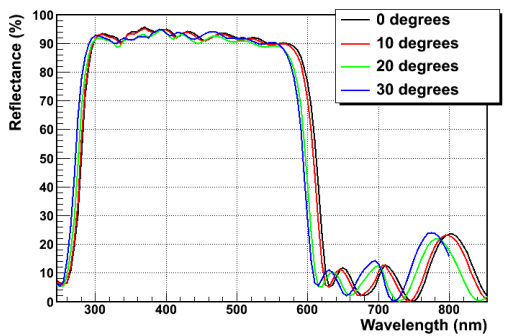}
\caption[Sim_Reflectance_TiO2-SiO2-HfO2_28layers]{\footnotesize Simulated reflectance vs. wavelength of the above-described 28-layers TiO$_{2}$-SiO$_{2}$-HfO$_{2}$ coating design, at different incident angles (0 degrees corresponds to normally incident light).}
\label{Sim_Reflectance_TiO2-SiO2-HfO2_28layers}
\end{figure}

The incident angle of Chererenkov light photons on the IACT mirror tiles changes according to their position on the IACT dish and to the $\gamma$-ray induced shower incoming direction. We report in Fig. \ref{Sim_Reflectance_TiO2-SiO2-HfO2_28layers} the reflectance curve of our 28-layers TiO$_{2}$-SiO$_{2}$-HfO$_{2}$ coating design at different incident angles (0, 10, 20 and 30 degrees) as obtained by the simulation software. As it can be seen, the average reflectance in WR$_{300-550}$ slightly decreases from 0 degrees (92.8$\%$) to 30 degrees (89.2$\%$). Nevertheless, it remains above $90\%$ for an incident angle up to 25 degrees ($92.1\%$), which corresponds to the maximum incident angle for a telescope whose focal length over dish diameter ratio is $\frac{f}{D}=1.2$, a typical value for IACTs. 

\section{Discussion and Conclusion}

As shown in Fig.~\ref{TiO2-SiO2_24layers}, according to our investigation it seems not possible to achieve a reflectance equal to $90\%$ in WR$_{300-550}$ with a multilayer coating made of TiO$_{2}$ and SiO$_{2}$ only, because of the poor TiO$_{2}$ transparency at wavelengths shorter than 350 nm. Therefore, any solution involving TiO$_{2}$ and SiO$_{2}$ without any third material will be suitable for longer wavelengths only.\\
The reflectance curves shown in Fig.~\ref{Reflectance_TiO2-SiO2-HfO2_28layers} prove that it is possible to overcome the poor TiO$_{2}$ transparency in the ultraviolet region by using few (4 in our case) additional alternating layers of HfO$_{2}$ and SiO$_{2}$.\\
As shown in Fig.~\ref{Sim_Reflectance_TiO2-SiO2-HfO2_28layers}, according to our simulations the incident angle dependence of the reflectance curve is very small up to 25 degrees. Usually interferometric devices are quite affected by changes of the incident angle, but the effect in our case is quite limited thanks to the large number of layers. The light incident angle range is expected to be encompassed between approximately 0 and 25 degrees for all the CTA 1-Mirror telescopes. 
In the case of the CTA Schwarzschild-Couder telescopes currently under design, the maximum incident angle is below 25 degrees for the primary mirror only, while for the secondary mirror the range of light incident angles is considerably broader. Therefore, any multilayer interferometric coating is most likely unfit for the secondary mirrors of such telescopes.\\


We want to stress that our coating solution can be easily applied for mirror mass production, and in particular for the CTA project. The final cost of our solution is expected to be larger compared to the Aluminum solution because of the increased deposition time. Thus, we expect that the cost increase would be similar to the H.E.S.S. dielectric mirrors, as written in sec.~\ref{Present_IACT_mirrors}. We want to point out also that it would be possible to obtain a similar result by using HfO$_{2}$ and SiO$_{2}$ only. In that case the number of layers would be even larger ($\sim50$ vs. 28) because HfO$_{2}$ refractive index is lower than TiO$_{2}$ one (1.96 vs. 2.42 for TiO$_{2}$ bulk material or 2.30 in our case, at $\lambda$=450~nm), thus leading to higher manufacturing time and cost.\\


As already stated above, like for any other multilayer interferometric coating, it is possible to tune the reflectance curve of our design (i.e. enlarging, shrinking, blue or red-shifting the bandwidth, increasing or decreasing the reflectance, and changing the optimized incident angle) by modifying the layer amount and thicknesses. Therefore, it is possible to shrink the reflectance region well below the strong OI NSB line at 557.7 nm, which is poorly suppressed in the current design, so to achieve a higher NSB suppression. On the other hand, any shrinkage of the reflection region has to be carefully evaluated, since it would turn out in a reduction of the detected Cherenkov light too (see Fig.~\ref{cherenkov_spectrum}).\\
It is also possible to apply our coating solution to other scientific and industrial purposes (e.g. mirrors for solar power plants). It might even be possible to adapt this solution to the IACT camera funnels but, due to their geometry which is far from a pure plane, their mass production could be cost challenging. Furthermore, due to the very large light incident angles ($\geq 60$~deg), the layer thicknesses should be completely revisited.\\

\section*{Acknowledgement}

The research leading to these results has received funding from the
European Union's Seventh Framework Program (FP7/2007-2013) under grant
agreement n° 262053.\\
This work has been partially funded by the BMBF/PT-DESY, grants 05A11VT1 and 05A10VTA.\\

We want also to thank the Max Planck Institut f\"ur Kernphysik (MPIK) Heidelberg, and specially Andreas F\"orster, for the help and the technological support. 


\end{document}